\documentclass[fleqn,usenatbib]{mnras}

\usepackage{newtxtext,newtxmath}

\usepackage[T1]{fontenc}

\DeclareRobustCommand{\VAN}[3]{#2}
\let\VANthebibliography\thebibliography
\def\thebibliography{\DeclareRobustCommand{\VAN}[3]{##3}\VANthebibliography}

\usepackage{graphicx}	
\usepackage{amsmath}	
\usepackage{subcaption}
\usepackage{makecell}
\usepackage{hyperref}
\usepackage{footmisc}
\usepackage{orcidlink}
\usepackage[export]{adjustbox}
\usepackage{upgreek}

\usepackage{academicons}
\definecolor{orcidlogocol}{HTML}{A6CE39}

\def\code#1{\texttt{#1}}

\newcommand{\RMunits}{rad$\,$m$^{-2}$}
\newcommand{\RMalpha}{rad$\,$m$^{-\alpha}$}
\newcommand{\RMalphathree}{rad$\,$m$^{-3}$}
\newcommand{\DMunits}{pc\,cm$^{-3}$}
\newcommand{\Qd}{Q_{\rm de-rot}}
\newcommand{\Ud}{U_{\rm de-rot}}
\hypersetup{
    colorlinks=true,
    linkcolor=blue,
    filecolor=magenta,      
    urlcolor=magenta,
    pdfpagemode=FullScreen,
    }
\title[Circular polarisation in FRB~20180301A]{Towards solving the origin of circular polarisation in FRB~20180301A}
\author[P. Uttarkar et al.]{\parbox{\textwidth}
{
Pavan Uttarkar,$^{1}$\orcidlink{0000-0002-2346-6853}\thanks{E-mail: puttarkar@swin.edu.au}
Ryan~M.~Shannon$^{1}$\orcidlink{0000-0002-7285-6348},
Marcus~E.~Lower$^{2}$\orcidlink{0000-0001-9208-0009},
Pravir~Kumar$^{3}$\orcidlink{0000-0003-1913-3092},
Danny~C.~Price$^{4, 5}$\orcidlink{0000-0003-2783-1608},\\
A.~T.~Deller$^{1}$\orcidlink{0000-0001-9434-3837},
K.~Gourdji$^{1}$\orcidlink{0000-0002-0152-1129}
}
\\ \\
$^{1}$Centre for Astrophysics and Supercomputing, Swinburne University of Technology, Hawthorn, VIC\\
$^{2}$Australia Telescope National Facility, CSIRO, Space and Astronomy, PO Box 76, Epping, NSW 1710, Australia\\
$^{3}$Department of Particle Physics and Astrophysics, Weizmann Institute of Science, 76100 Rehovot, Israel\\
$^{4}$Square Kilometre Array Observatory (SKAO)\\
$^{5}$International Centre for Radio Astronomy Research (ICRAR)\\
}

\date{Accepted XXX. Received YYY; in original form ZZZ}

\pubyear{2015}

\begin{document}
\label{firstpage}
\pagerange{\pageref{firstpage}--\pageref{lastpage}}
\maketitle

\begin{abstract}
Fast Radio Bursts (FRBs) are short-timescale transients of extragalactic origin. The number of detected FRBs has grown dramatically since their serendipitous discovery from archival data. Some FRBs have also been seen to repeat. The polarimetric properties of repeating FRBs show diverse behaviour and, at times, extreme polarimetric morphology, suggesting a complex magneto-ionic circumburst environment for this class of FRB. The polarimetric properties such as circular polarisation behaviour of FRBs are crucial for understanding their surrounding magnetic-ionic environment. The circular polarisation previously observed in some of the repeating FRB sources has been attributed to propagation effects such as generalised Faraday rotation (GFR), where conversion from linear to circular polarisation occurs due to the non-circular modes of transmission in relativistic plasma.
The discovery burst from the repeating FRB~20180301A showed significant frequency-dependent circular polarisation behaviour, which was initially speculated to be instrumental due to a sidelobe detection.
Here we revisit the properties given the subsequent interferometric localisation of the burst, which indicates that the burst was detected in the primary beam of the Parkes/Murriyang 20-cm multibeam receiver. We develop a Bayesian Stokes-Q, U, and V fit method to model the GFR effect, which is independent of the total polarised flux parameter. Using the GFR model we show that the rotation measure (RM) estimated is two orders of magnitude smaller and opposite sign ($\sim$28 \RMunits) than the previously reported value. We interpret the implication of the circular polarisation on its local magnetic environment and reinterpret its long-term temporal evolution in RM. 
\end{abstract}




\section{Introduction}

Fast radio bursts (FRBs) are microsecond to millisecond duration dispersed bursts of radio emission of extragalactic origin. The first FRB was reported by \cite{Lorimer} in a 20-cm multibeam receiver archival dataset from the Parkes/Murriyang radio telescope. Since the first reported burst, the number of  FRBs discovered has grown to $>$2000 FRBs to date \citep{Xu_blinkverse}. A large fraction of the new bursts detected have been reported by instruments commissioned during the last decade, such as the Australian Square Kilometer Array Pathfinder \citep[ASKAP;][]{Bannister_2017}, Canadian Hydrogen Intensity Mapping Experiment \citep[CHIME;][]{CHIME_catalog}, Deep Synoptic Array \citep[DSA;][]{DSA_110}. Even with a rapid increase in the number of detected FRBs, the progenitor of FRBs remains elusive, with much of the physics (e.g., emission mechanism, energy distribution, surrounding magnetic field) yet to be fully explained.

Among the sample set of FRBs discovered, $\sim$65 FRBs \citep[$\sim$2\% of the population;][]{CHIME_catalog} have been observed to repeat. Whether all FRBs repeat (eventually) is  unclear. Soon after the discovery of a population of repeating FRBs, it became evident that there were differences in the spectro-temporal properties (e.g., pulse morphology, spectral extent) of repeaters and (apparent) non-repeaters \citep{CHIME_catalog}, with the linear frequency drift, or "sad trombone" being a distinct spectral behaviour seen in some of the repeating FRB sources \citep[e.g.,][]{Hessels,Luo_2020,CHIME_catalog}. Additionally,  smaller spectral extents are observed in repeating FRB sources compared to non-repeaters, with the spectral and temporal widths in repeating FRB sources found to be a function of frequency \citep{Bethapudi}.
Further, some repeaters such as FRB 20180916B, have been shown to have burst activity windows that are frequency-dependent. \citep[e.g.,][]{Pleunis_chromatic, Bethapudi}.

In addition to the spectro-temporal characteristics, the polarisation properties of FRBs provide crucial inferences on the FRB emission mechanism, circumburst media, and potential progenitor. A linearly polarised radio wave when travelling through a dispersive, cold magnetised plasma such as interstellar medium (ISM), or intergalactic medium (IGM) can induce typical nanosecond scale delay between the left and right handed circular polarisation modes, resulting in Faraday rotation (FR).
The effect of FR can be quantitatively described using the Rotation Measure (RM), a measure of the quadratic variation in linear polarisation position angle (PPA) with the frequency. In general, FRBs have been observed to display a 
plethora of additional polarisation properties, such as the PPA swing seen in FRB~20180301A \citep{Luo_2020} relative to a flat PPA seen in most other FRBs \citep[e.g.,][]{Bannister_2019, Day_2020, Pravir_190711}. Further, a significantly large variation in linear polarisation fraction has been observed between different one-off bursts, and also from an active repeating source \cite[e.g.,][]{Day_2020, Pravir_20201124A_MNRAS}. In addition to the conventional FR effect, which is caused by the circularly polarised natural transmission modes, some repeaters such as FRB 20201124A have shown to have significant frequency-dependent circular polarisation fraction \citep{Pravir_20201124A_MNRAS}. The frequency-dependent circular polarisation has been attributed to Faraday conversion or generalised Faraday rotation (GFR) effect \citep{Pravir_20201124A}. 
GFR can induce circular polarisation by the conversion of the linear Stokes components to circular polarisation. 
The propagation of polarised radiation through a medium having an elliptical or linear transmission mode results in such conversion \citep{Kennett}.  
Like conventional Faraday rotation, GFR is a chromatic process. Hence, it will result in strong frequency dependence for the circular polarisation fraction. 

Previous investigations of pulsars \citep[e.g.,][]{Melrose_2003, Melrose_2004}  point to three main causes of circular polarisation: (a) a process intrinsic to the emission; (b) a favorable natural transmission mode; (c) 
propagation effects (e.g., in relativistic plasma). Additionally, some studies such as \cite{Gruzinov_2019} have also proposed the generation of circular polarisation through conventional cold plasma - without relativistic electrons - due to the magnetic field reversals along the line of sight. 

There has also been a rapid increase in repeating FRBs showing extreme polarisation behaviour (e.g., 20121102A, 20190520B, and 20201124A). Two FRBs, 20190520B and 20180301A have been observed to undergo a change in the sign of the RM, suggesting a reversal in the parallel magnetic field component \citep{Anna_Thomas_190520B, Pravir_2023}. 
Additionally, some repeaters have been seen to have spectral depolarisation towards lower frequencies, suggesting a multipath propagation effect \citep{Feng_2022}. 
In general, the magnitude of circular polarisation observed in repeating FRBs is relatively smaller than in non-repeating FRB sources. Some sub-components of (apparent) non-repeaters, such as FRB 20190611B, have been seen to have a high circular polarisation fraction \citep[$\sim$57\%;][]{Day_2020}. 
The circular polarisation in FRBs can be used as a unique probe to understand the progenitor environment and its local complex magneto-ionic environment if the Faraday conversion occurs in its circumburst medium.  

The circular polarisation observed in several repeating FRB sources (e.g., FRB 20201124A, 20220808A) has been interpreted as being caused by GFR. FRB 20201124A was one of the first repeaters to have been observed to have significant and notable frequency dependent circular polarisation \citep[$\sim$57\%;][]{Pravir_20201124A_MNRAS}.
While FRB~20180301A was the first repeater to have been seen to have a significant frequency-dependent circular polarisation; it was initially speculated that the circular polarisation was induced due to instrumental leakage due to a sidelobe detection \citep{Price_2018}. However, the milli-arcsecond localisation by \cite{Bhandari_2022} has provided precise localisation information which can now be leveraged 
to study the Parkes/Murriyang discovery burst spectro-polarimetric characteristics.  

In this paper, we revisit the spectro-polarimetric properties of FRB~20180301A using GFR modelling. We summarise the data used in this analysis, and the GFR modelling used to explain the circular polarisation behaviour in Section \ref{sec:Methods}. We discuss the results and their implications from the GFR modelling in Sections \ref{sec:Results} and \ref{sec:Discussion}, respectively. We conclude our analysis in Section \ref{sec:Conclusisons}.


\begin{figure*}
         \includegraphics[width=0.6\textwidth]{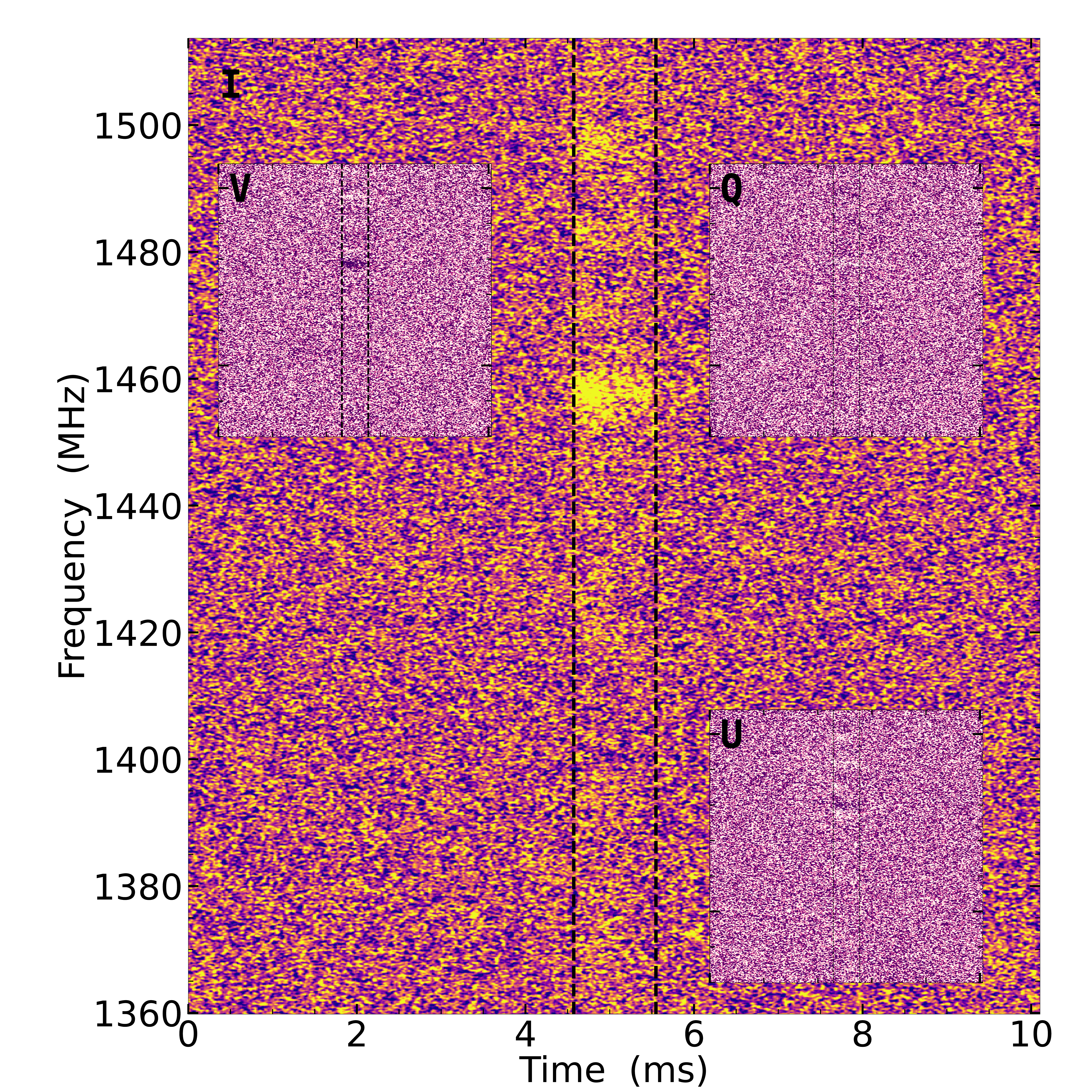}
         \caption{The dynamic Stokes spectra of FRB~20180301A. The dynamic Stokes I, Q, U, and V spectra are shown in Panels I, Q, U, and V, respectively. The lighter regions show higher flux relative to the darker regions in the spectra. The dotted line shows the  integrated region used to generate the time-averaged spectra for GFR modelling. The frequency dependent behaviour of Stokes-V can be seen in the banding across frequency in Panel V. The dynamic spectra are smoothed with a Gaussian filter having a frequency and temporal width of 0.11 MHz and 15.24$\upmu$s, respectively.}
         \label{fig:spectra_180301A}
\end{figure*}

\section{Methods}
\label{sec:Methods}
\subsection{Data description and preparation}

We use data presented in \cite{Price_2018} for our analysis. Here we briefly summarise the properties of the data.   FRB~20180301A was discovered using the 21-cm multibeam receiver on the Parkes/Murriyang radio telescope \citep{Staveley-Smith}, using the Berkeley–Parkes–Swinburne Recorder (BPSR) system in a Breakthrough Listen (BL) observation \citep{Price_2018}. Repeat bursts from FRB~20180301A were reported using the Five Hundred Meter Aperture Radio Telescope (FAST)  by \cite{Luo_2020}, confirming the source to be a repeating FRB. The initial burst was reported to have an RM of -3156 \RMunits \citep{Price_2018} and a dispersion measure (DM) of 522 \DMunits. The dynamic Stokes spectrum from the multibeam receiver is shown in Figure \ref{fig:spectra_180301A}. We use methods described in \cite{Price_2018} to flux calibrate the data on the existing polarisation calibrated dataset between the frequencies 1359.8 MHz and 1513.7 MHz, for the 154 MHz bandwidth over which the the FRB emission is apparent. We do not include the lower 154 MHz in our analysis, where there is no detectable burst emission.
Following \cite{Price_2018}, we use a Gaussian filter with a frequency width of 0.109 MHz and a time width of $\sim$60 $\upmu$sec to smooth the spectra. 

\begin{figure}
         \includegraphics[width=0.47\textwidth]{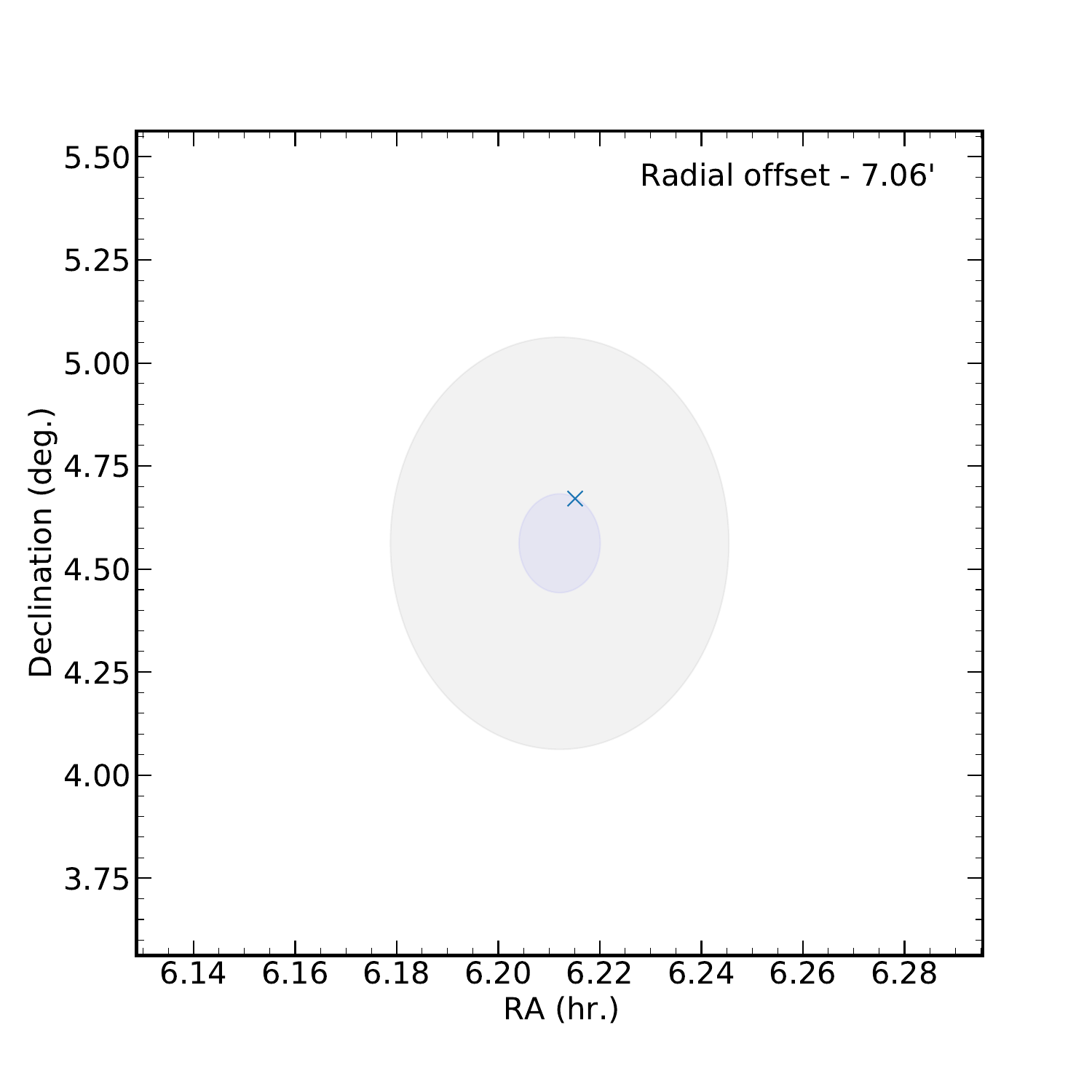}
         \caption{Localisation of FRB~20180301A. The VLA localisation of a repeat burst reported by \protect \cite{Bhandari_2022} is shown by the blue cross. The FWHP of beam 3 relative to the localisation is shown by the blue shaded region. The grey shaded region shows the extent of a single beam. FRB~20180301A was radially offset by $\sim$7' from the beam center. The localisation information of the burst and the Parkes beam pointing information suggest a beam 3 primary lobe detection of the burst.}
         \label{fig:localisation}
\end{figure}

The burst was discovered in beam $3$ of the multibeam receiver. The lack of precise localisation after the initial discovery led to the speculation that the burst was detected in a sidelobe \citep{Price_2018}. Hence, to explain frequency-dependent circular polarisation, polarisation leakage due to the beam 3 sidelobe detection was considered as one of the likely scenarios. However, the subsequent $\sim$milli-arcsec localisation by \cite{Bhandari_2022}, and the pointing position of the Parkes/Murriyang observation confirm a primary-lobe detection of the burst. Considering a full-width half-power (FWHP) beamwidth of $\sim$14.5' \citep{Staveley-Smith} for beam 3, the multibeam detection shows a radial offset of $\sim$7'. Figure \ref{fig:localisation} shows the FWHP of beam 3 in the blue-shaded region and the full-width (FW) of beam 1 in the grey-shaded region to show the general extent of the FW of individual beams. Additionally, the cross-polarisation leakage analysis by \cite{Carozzi} shows that the leakage of linear polarisation into circular polarisation is smaller than 15-20 dB (a factor of 30-100), indicating low polarisation leakage in the multibeam receiver. Hence, the frequency-dependent circular polarisation structure in the burst is likely intrinsic to the source and not anthropogenic or an instrumental artifact, as speculated by \cite{Price_2018}. 

We model the spectral behaviour of the polarisation structure in the burst using a phenomenological GFR model described in \cite{Lower_2021}. We extend this model to include the modelling of RM from the conventional FR effect. We use this FR-GFR model to investigate the polarised behaviour of the burst. To verify the model, we use data from the Parkes/Murriyang multibeam receiver observations of known pulsars J1644$-$4559 and J0835$-$4510, which have reported measurements of RM of -626.9\footnote{\url{https://www.atnf.csiro.au/research/pulsar/psrcat/}\label{fn:psrcat}} and 31.38\footref{fn:psrcat}, respectively. 
The data for pulsars J0833$-$4559 and J0835$-$4510 were taken in $2018$ and recorded using the Parkes Digital Filterbank Backend Mark-4.  
The centre frequency for both observations is 1369 MHz, with a bandwidth of 256 MHz, with a frequency resolution of 125 KHz. We calibrate the pulsar data for flux and polarisation using a standard Parkes flux calibrator and a noise diode source for polarisation calibration using \code{psrchive}.

\begin{figure}
         \includegraphics[width=0.47\textwidth]{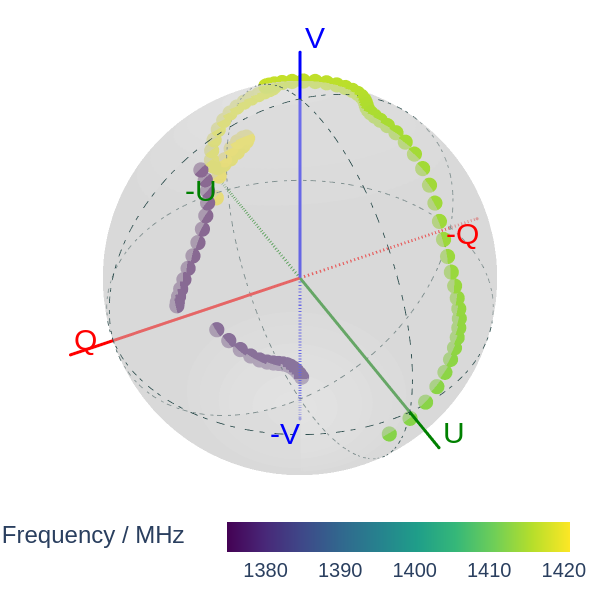}
         \caption{Poincare sphere representation of the spectropolametric properties of FRB~20180301A. The data are time averaged and smoothed using a Gaussian filter with a frequency width of 0.109 MHz and a time constant of $\sim$60 $\upmu$sec. The colour map shows the time averaged measurements at different frequencies. The polarisation information is shown for the data points having an S/N greater than 4.5 in individual frequency channels of 0.109 MHz bandwidth.}
         \label{fig:Poincare}
\end{figure}

The data projected on the Poincare sphere can provide useful visualisation of the polarimetry, in particular for GFR analysis. We show the projection of the RM uncorrected Stokes components on the Poincare sphere across frequency, with measurements having a fractional polarised intensity greater than ~5\% in Figure \ref{fig:Poincare}.
The polarisation vector of a purely FR-dominated polarisation source does not transverse different latitudes on the Poincare sphere. However, a frequency-dependence  of Stokes-V could result in the polarisation vector moving to different latitudes. 
The projection of the polarisation vector on the Poincare sphere for FRB~20180301A shows a clear frequency-dependent structure across the frequency band, suggesting linear to circular polarisation conversion across  frequency.  

\subsection{Faraday rotation - Generalised Faraday rotation (FR-GFR) model}

\subsubsection{Modelling Faraday rotation}
We use the FR-GFR model to estimate both RM and GRM for the data. In the FR-GFR model we calculate the PPA for FR and GFR cases simultaneously. The PPA for the case of the conventional FR model (index $\alpha$=2) is given by 

\begin{equation}
    \psi = \psi_{o} + \rm RM (\lambda^2 - \lambda_{o}^2),
    \label{eq:PPA_quadratic}
\end{equation}
where RM quantifies the magnitude of phase change induced by the intervening cold magnetised plasma, $\lambda$ is the wavelength, $\lambda_{o}$ is the wavelength associated with a reference frequency.
The RM can be related to physical quantatites using 
\begin{equation}
    {\rm RM} = \frac{e^3}{2\pi m_e^2 c^4}\int_{d}^{0}\frac{n_e B_{||}}{(1+z)^2} dl,
	\label{eq:RM_calulation}
\end{equation}
where $e$ is the electron charge, $z$ is the redshift of the electrons, $m_e$ is the mass of the electron, $c$ is the speed of light, $n_e$ is the free electron density, and $B_{||}$ of the magnetic field component parallel to the line of sight.

\subsubsection{Modelling non-circular modes of transmission}
In addition to the conventional FR, where the natural transmission modes are circularly polarised, conversion from linear to circular polarisation can be induced if the modes of transmission of the radio waves are either non-circular (elliptical) or completely linear. The non-circular modes of transmission can also lead to a non-$\lambda^2$ dependency of PPA, with the chromaticity dependent on the properties of the plasma and underlying theoretical assumptions. These non-circular modes can be modelled by using an arbitrary spectral scaling index $\alpha$. Hence, Equation \ref{eq:PPA_quadratic} can be generalised for the case of GFR to be
\begin{equation}
\Psi(\lambda) = \Psi_{o} + \rm GRM (\lambda^\alpha - \lambda_{o}^\alpha), 
\end{equation}
where $\Psi_{o}$ is the intrinsic PPA, GRM is generalised rotation measure (GRM) an analogue of RM for the non-circular modes of transmission. 

The polarisation vector due to the effects of cold plasma and GFR it modelled to be
\begin{equation}
    \bf{P}_{\rm GFR} = \bf{R}{\psi} \bf{R}_{\theta \phi} \bf{P}(\lambda),
    \label{eq:Circular_model}
\end{equation}
where \textbf{R}$_{\psi}$ is the FR-induced shift in the polarisation vector, with a fixed latitude on the Poincare sphere, \textbf{R}$_{\theta \phi}$ is the rotation matrix describing the arbitrary rotation of the polarisation vector on Poincare sphere over different latitudes (a characteristic of the GFR; see \cite{Lower_2021} for a detailed discussion) and \textbf{P}($\lambda$) is the Stokes vector, which is a function of PPA and the elipticity angle (EA). 

The rotation matrices \textbf{R}$_{\phi}$, \textbf{R}$_{\theta \phi}$, and \textbf{P}($\lambda$), which are used to model the angle of polarisation vector on a Poincare sphere can be described by
\begin{equation}
\textbf{R}_{\rm \psi} = 
\begin{pmatrix}
\cos 2\psi & -\sin 2\psi & 0\\
\sin 2\psi & \cos 2\psi & 0\\
0 & 0 & 1\\
\end{pmatrix},
\end{equation}
\begin{equation}
\textbf{R}_{\rm \theta \phi} = 
\begin{pmatrix}
\cos(\theta) \cos(\phi) & -\cos(\theta)\sin(\phi) & \sin(\theta)\\
\sin(\phi) & \cos(\phi) & 0\\
-\sin(\theta)\cos(\phi) & \sin(\theta)\sin(\phi) & \cos(\theta)\\
\end{pmatrix},
\end{equation}
and
\begin{equation}    
\bf{P}(\lambda) = P
\begin{pmatrix}
\cos(2\Psi) \cos(2\chi) \\
\sin(2\Psi) \cos(2\chi) \\
\sin(2\chi)\\
\end{pmatrix},
\label{eq:Stokes}
\end{equation}
where $\theta$ describes the angle of the axis about which the polarisation vector rotates, $\phi$ is the rotation about the Stokes-Q axis on the Poincare sphere, $\Psi_o$ is the intrinsic polarisation angle, and $\chi$ is the EA describing the angle made by the polarisation vector with respect to the Stokes-V axis, given by
\begin{equation}  
\chi = \frac{1}{2} \arctan\left(\frac{V}{\sqrt{Q^2+U^2}}\right). 
\end{equation}

\subsubsection{Bayesian estimation of the GFR parameters}

We perform a Bayesian parameter inference using \code{bilby}\,\citep{Ashton} with the \code{DYNESTY} nested sampler to infer the posterior probability distribution for our models. We expand the method described in \cite{Bannister_2019} to infer Faraday Rotation to include GFR-induced circular polarisation. Assuming the noise in the data can be described by a Gaussian distribution (valid if the thermal noise dominates the data), we can infer the posterior probability distribution for the GFR to be
\begin{equation}
     \textit{L}(P(\lambda)| P_M(\lambda) , \sigma) =  \prod_i^{N} \frac{1}{\sqrt{2\pi\sigma^2}}\,\,\exp\left(-\frac{(P_i(\lambda)-P_{M,i}(\lambda))^2}{2\sigma^2}\right),
    \label{eq:posterior}
\end{equation}

where $P(\lambda)$ is the polarisation vector of the data consisting of all the Stokes components, $P_M(\lambda)$ is the modeled polarisation vector with all the Stokes parameters, $\sigma$ is the standard deviation of the Stokes spectrum, and $\lambda$ is the wavelength. The posterior in the basis is
\begin{equation}
\begin{split}
     &\textit{L}(P(\lambda)| P_M(\lambda) , \sigma) =
     \prod_i^{N} \frac{1}{(2\pi)^{3/2}\sigma^3}\\
     &\int_{0}^{\infty}\exp\left[-\frac{(Q-Q_m)^2}{2\sigma^2}\right]
     \exp\left[-\frac{(U-U_m)^2}{2\sigma^2}\right]
     \exp\left[-\frac{(V-V_m)^2}{2\sigma^2}\right] \,dP,
\end{split}
\label{eq:postrior_model}
\end{equation}
where Q, U, and V are the Stokes components from the data, and $Q_{m}$, $U_{m}$, and $V_{m}$ the modelled  Stokes components.

Following \cite{Bannister_2019}, we marginalise the above Equation over the total polarisation fraction $P$ to make the estimation more robust to changes in the polarised flux in any of the frequency channels. 
We modify the integration intervals from 0 to $\infty$ from -$\infty$ to $\infty$ used in \cite{Bannister_2019}. We modify the integration interval as the latter leads to a bi-model distribution in the likelihood function for PPA and EA.  We consider two plausible scenarios: either the noise in all the Stokes components is identical, or they are all different, which we term, cases (a) and (b), respectively.



For the case (a), the marginal posterior distribution function is  
\begin{equation}
\begin{split}
& \textit{L}(P(\lambda)| P_M(\lambda) , \sigma) = \prod_i^{N} \frac{1}{2\pi\sigma^2} \\
&\exp\left[{-\frac{1}{2\sigma^2}\left( (Q^2+U^2+V^2) - (QQ_{m}+UU_{m}+VV_{m})^2\right)}\right]\\
& \rm \left[erf\left({\frac{\sigma}{\sqrt{2}} \left(QQ_{m}+UU_{m}+VV_{m}\right)}\right)+1\right],
\end{split}
\label{eq:posterior_final}
\end{equation}
where $Q_{m}$, $U_{m}$, $V_{m}$ are the modeled Stokes vectors from Equation \ref{eq:Stokes}.

For case (b), the marginal posterior distribution is
\begin{equation}
\begin{split}    
&   \textit{L}(P(\lambda)| P_M(\lambda) , \sigma)  = 
    \frac{1}{4\pi\sqrt{C_2}} \exp{\Biggr[\frac{C_1}{2N_T}\Biggr]} \\ 
&   \exp{\Biggr[\frac{C_3^2}{2C_2N_T}\Biggr]} 
   \rm \Biggr[erf\left(\frac{C_3}{\sqrt{2N_TC_2}})\right) + 1\Biggr],
\end{split}
\label{eq:likelyhood_case_b}
\end{equation}
where the factors $N_T$, $N_1$, $N_2$, $N_3$, $C_1$, $C_2$ and $C_3$ are 
\begin{equation}
    \begin{split}
    &   N_T = \sigma_Q^2\sigma_U^2\sigma_V^2,\\
    &   N_1 = \sigma_U^2\sigma_V^2, \\
    &   N_2 = \sigma_Q^2\sigma_V^2, \\
    &   N_3 = \sigma_Q^2\sigma_U^2, \\
    &   C_1 = Q^2N_1 + U^2N_2 + V^2N_3,\\
    &   C_2 = Q_m^2N1 + U_m^2N_2 + V_m^2N_3, \textnormal{and}\\
    &   C_3 = QQ_mN_1 + UU_mN_2 + VV_mN_3.\\
    \end{split}
\end{equation}

\begin{figure}
         \includegraphics[width=0.5\textwidth]{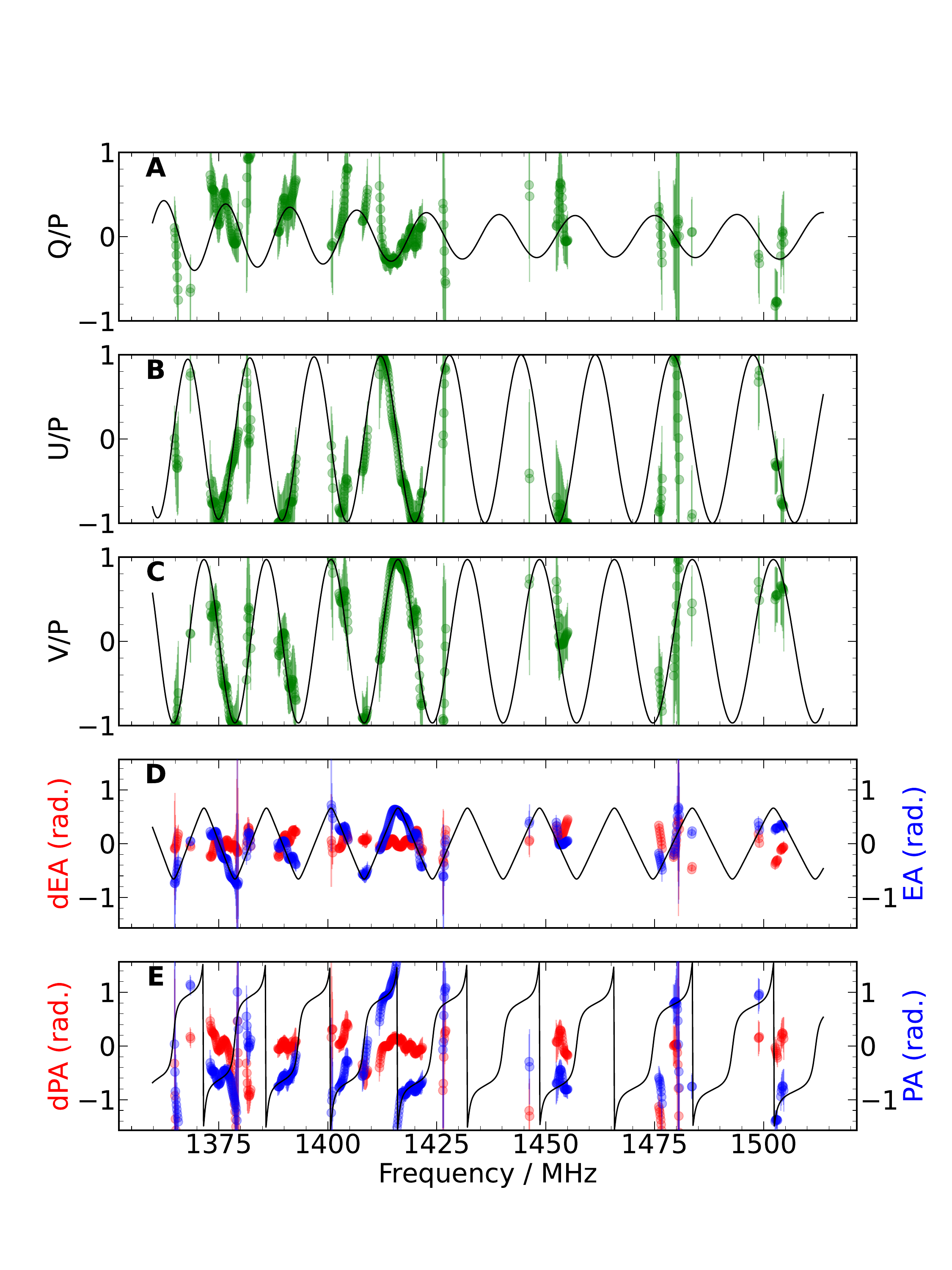}
         \caption{Maximum likelihood GFR model for FRB~20180301A.  The QUV fit to the FRB~20180301A from the FR-GFR model for Stokes Q, U, and V fit are shown in Panels A, B, and C, respectively. The data from the burst are shown in green coloured points, and the solid black line shows the model fit to the data. The EA and PA angles derived from the data are shown as blue points in the Panels D and E, respectively. The model fit for the EA and PA angles are shown in black solid lines. The residuals after subtracting the data from the model are shown in red points for EA and PA angles in Panels D and E.}
         \label{fig:180301A_QUV_fit}
\end{figure}

We sample the uncertainties in the Stokes spectrum in the parameter estimation and calculate the uncertainties in the PPA and the EA using 

\begin{equation}
    \sigma_{\psi} = \sqrt{\frac{\Qd^2 \sigma^2 + \Ud^2  \sigma^2 }{4(\Qd^2 + \Ud^2)}}, 
\end{equation}
and
\begin{equation}
    \sigma_{\chi} = \frac{\sqrt{(V\Qd)^2\sigma^2 + (V\Ud)^2\sigma^2 (\Qd^2+V^2)^2\sigma^2}}{4(\Qd^2+\Ud^2+V^2)\sqrt{\Qd^2+\Ud^2}}.
\end{equation}

Bayesian parameter estimation is undertaken on a complete dataset without excising any frequency channels since the posterior probability distribution function described by Equation \ref{eq:posterior_final} is agnostic to the total polarisation fraction of the burst. Hence, it is immune to any spurious change in total polarised amplitude in the channels, making it more robust relative to the GFR likelihood function described in \cite{Lower_2021}. We use uniform priors for all the input parameters in our estimation.

\section{Results}
\label{sec:Results}

\subsection{Model verification on known sources}

We verify the FR-GFR model on known pulsars from archival observations\footnote{\url{https://data.csiro.au/}} from the Parkes/Murriyang multibeam receiver using the model described by Equation \ref{eq:Circular_model} and the likelihood function described by Equation \ref{eq:likelyhood_case_b}. The posterior probability distributions for J1644-4559 and J0835-4510 are shown in Figure \ref{fig:pulsar_posteiors}. We show the model fit for Stokes Q, U, and V from the estimated parameters in Figure \ref{fig:pulsar_QUV_fit}. The FR-GFR model estimates the RM for J1644-4559 and J0835-4510 to be -626.31$_{-1.2}^{+1.2}$ \RMunits and 48.36$_{-0.9}^{+0.9}$ \RMunits, respectively. The estimated parameters for the FR-GFR model, \code{rmfit}, and FR model are listed in Table \ref{tab:comparision_table}. 
The FR-GFR polarisation parameters were estimated by averaging 154 and 174 bins over the on-pulse region for J1644-4559 and J0835-4510, respectively. The RM for the pulsars J1644-4559 and J0835-4510 were computed by \code{psrchive}$-$\code{rmfit} using a resolution of 0.5 \RMunits over the range -1000 to 0 \RMunits\, and 0 to 50 \RMunits\,, respectively.
The posterior probability distribution for parameters GRM and $\alpha$ return one of the prior boundaries for the case of J1644-4559, indicating a near constant Stokes-V across the observation frequency band. Similarly, the index $\alpha$ for J0835-4510 also returns one of the prior boundaries, indicating a constant Stokes-V across the frequency band for J0835-4510, even though the  GRM is inferred to be 71$_{-15}^{+20}$. 
We further verify the FR-GFR model on magnetar XTE J1810-197 from \cite{Lower_2023} to confirm the recovery of the expected GRM and $\alpha$ values by comparing the results to \cite{Lower_2023}. We show the GRM and $\alpha$ and values of our fits in Figure \ref{fig:J1810_verififcation} in the appendix.  The results are in broad agreement with those derived  in \cite{Lower_2023}.  We attribute the difference in the recovered GRM and $\alpha$ values  to be due to the simultaneous RM measurement in our approach. 

\begin{table*}
	\begin{center}
		\begin{tabular}{ c c c c c c c c c }
			\hline
			Pulsar & method & \thead{RM \\ (rad m$^{-2}$)} & \thead{GRM \\ (rad m$^{-\alpha}$)} & \thead{$\psi$ \\ (deg)} & \thead{$\chi$ \\ (deg)} & \thead{$\phi$ \\ (deg)} & \thead{$\alpha$ \\ } & \thead{$\theta$ \\ (deg)} \\
			\hline
                \\
			J1644-4559 & \code{FR-GFR} & -626.3$_{-1.2}^{+1.2}$ & < 433.49 & 54.6$_{-0.3}^{+0.3}$ & 42.0$_{-0.1}^{+0.1}$ & -62.2$_{-3.4}^{+3.2}$ & < 1.98 & 73.6$_{-0.2}^{+0.2}$ \\
			\\
			 & \code{rmfit} & -621.0$_{-2.3}^{+2.3}$ & - & - & - & - & - & - \\
			\\
			 & \code{FR} & -623.1$_{-1.3}^{+1.3}$ & - & 54.0$_{-0.3}^{+0.3}$ & 8.3$_{-0.1}^{+0.1}$ & - & - & - \\
                \\
			\hline
		      \\	
                J0835-4510 & \code{FR-GFR} & 48.4$_{-0.9}^{+0.9}$ & 71$_{-15}^{+20}$ & 82.0$_{-0.2}^{+0.2}$ & -42.4$_{-0.1}^{+0.1}$ & -86.7$_{-5.9}^{+5.9}$ & 1.2$_{-0.7}^{+0.3}$ & 78.8$_{-0.3}^{+0.3}$ \\
			\\
                & \code{rmfit} & 46.8$_{-0.01}^{+0.01}$ & - & - & - & - & - & - \\
                \\
			 & \code{FR} & 47.0$_{-0.6}^{+0.5}$ & - & -7.0$_{-0.1}^{+0.1}$ & -5.7$_{-0.1}^{+0.1}$ & - & - & - \\
			\\
                \\
			 
			\hline

		\end{tabular}
	\end{center}
	\caption{Comparison of Faraday conversion measurements for PSRs J0835-4510 and J1644-4559. The table lists estimated polarisation parameters from FR-GFR, \code{rmfit}, and FR methods.} 
	\label{tab:comparision_table}
\end{table*}

\subsection{Modelling GFR in FRB~20180301A}
In contrast to the two test pulsars described in the previous section, FRB~20180301A returns constrained posterior parameters. The posterior probability distribution of FRB~20180301A for the FR-GFR mode is shown in Figure \ref{fig:180301A_posteiors}. The RM and GRM value estimated using the FR-GFR model is 27.7$^{+5.22}_{-5.19}$ \RMunits and 4351.7$^{+340.40}_{-313.61}$ \RMalpha, respectively. The RM derived from the FR-GFR model is significantly smaller in magnitude than the previous RM estimation of -3163 \RMunits \citep{Price_2018}. The RM derived by \cite{Price_2018} relied on QU fit without considering the Stokes-V fit to the data. Moreover, the parameter $\theta$, which described the modes of transmission converges to $\sim$104$^{\circ}$, which indicates non-circular and nearly elliptical mode of transmission. The converged model parameters RM, GRM, $\psi$, $\chi$, $\phi$, $\alpha$ and $\theta$ for FRB~20180301A are listed in Table \ref{tab:180301A_tab}. The model fit to the data is shown in Figure \ref{fig:180301A_QUV_fit}. The index $\alpha$ converges to a $2.29 \pm 0.07$. The EA and PA angles calculated from the polarisation data across the frequency are shown in panels D and E in Figure \ref{fig:180301A_QUV_fit}. Additionally, the FR-GFR model has a log$_{10}$ Bayes evidence of 1389.6 relative to the FR-only model, indicating a strong preference \citep{Trotta_2008} to FR-GFR model. 

\begin{table}
	\begin{center}
		\begin{tabular}{c c}
			\hline
                \\
                Source & FRB 20180301A \\
                \\
			\hline
			\thead{RM  (rad m$^{-2}$)} & 27.7$_{-5.1}^{+5.2}$ \\
			\thead{GRM  (rad m$^{-\alpha}$)} & 4351.7$_{-313.6}^{+340.4}$ \\
			\thead{$\psi$  (deg.)} & -87.3$_{-0.7}^{+0.8}$ \\
			\thead{$\chi$  (deg.)} & -0.1$_{-0.4}^{+0.4}$ \\
			\thead{$\phi$  (deg.)} & 76.3$_{-1.0}^{+1.1}$ \\
			\thead{$\alpha$} & 2.3$_{-0.07}^{+0.07}$ \\
			\thead{$\theta$  (deg.)} & 104.2$_{-0.9}^{+1.0}$ \\
			\hline
		\end{tabular}
	\end{center}
	\caption{The converged GFR parameters for FRB 20180301A. The listed parameters are derived from the FR-GFR model using the likelihood Equation \protect\ref{eq:likelyhood_case_b}. The analysis was performed on a data set having frequency resolution of 0.109 MHz. The Stokes spectra were averaged using a simple boxcar.}
	\label{tab:180301A_tab}
\end{table}

\section{Discussion}
\label{sec:Discussion}

\subsection{Rationale for the polarimetric reanalysis of FRB~20180301A}
\begin{enumerate}
    \item[1.] The subsequent localisation from \citep{Bhandari_2022} and the beam pointing information from BL detection of FRB~20180301A \cite{Price_2018} shows that the detection of the burst was in the FWHP of beam 3 of the multibeam receiver. Hence, the circular polarisation behaviour in the burst is  not likely the result of instrumental leakage.
    \item[2.] Follow-up observation from the Parkes/Murriyang ultra wide band low (UWL) and FAST \citep{Luo_2020} has shown that FRB~20180301A has complex polarisation behaviour e.g., a reported switch in the sign of RM, tentative evidence of anti-correlation in the variation of DM and RM \citep{Pravir_2023}.
    \item[3.] Since the first discovery of the burst, other repeaters have been shown to have significant circular polarisation in some of their bursts \citep{Feng_20121102_20190520, Feng_220912}. Hence, it is  plausible that the circular polarisation is astrophysical. Such circular polarisation behaviour in their repetitions can be leveraged to study the surrounding media, and answer some of the outstanding questions related to the class of FRBs. 
\end{enumerate}

\subsection{Implications of the GFR modelling on FRB~20180301A}
Bursts from FRB~20180301A have been observed to have complex spectral,  temporal, and polarimetric properties, such as the secular variations in the RM in the previous Parkes and FAST observations \citep[e.g.,][]{Luo_2020, Pravir_20201124A_MNRAS}.
Among the known repeater sample, only $\sim$3 FRBs (FRBs 20190520B, 20201124A, and 20221912A) have been observed to have a relatively high circular polarisation fraction \citep{Pravir_20201124A_MNRAS, Feng_20121102_20190520}. In addition to a high circular polarisation fraction, FRB 20201124A was seen to have significant frequency-dependent circular polarisation  \citep{Pravir_20201124A_MNRAS}, suggesting linear to circular polarisation conversion due to relativistic plasma \citep{Pravir_20201124A}. Similar frequency-dependent circular polarisation behaviour was first observed in FRB~20180301A \citep{Price_2018}. However, because of an uncertain localisation of the FRB, it was speculated to be due to the instrumental polarisation leakage, potentially because of a sidelobe detection \citep{Price_2018}. The available localisation information points to the origin of the circular polarisation to be non-instrumental. Hence, the polarisation modelled using the GFR explains the origin of the Stokes-V emission in the source. 

An extensive follow-up campaign from Parkes/Murriyang UWL and FAST has shown that the FRB~20180301A has undergone significant RM evolution since its discovery \citep{Luo_2020, Pravir_2023}, including an apparent switch in sign in RM \citep{Pravir_2023}. The FRB 20190520B was the first repeater to have been seen to have such reversal in the sign of RM \cite{Anna_Thomas_190520B}, with a flip from an extreme RM of $\sim$-10$^4$ \RMunits to $\sim$10$^4$ \RMunits. In general, such RM reversal due to the reorientation of the magnetic field parallel component has been speculated to be due to various progenitor models, such as a magentar-Be star binary system \citep{Wang_model} (such as the B1259-63 binary system \cite{B1259}), or even due to the FRB progenitor around an intermediate mass black-hole \citep{Anna_Thomas_190520B}. In the case of a binary system, as suggested by \cite{Wang_model}, the RM evolution should be periodic, which is yet to be observed. Further, complex polarisation behaviour has been observed from active repeaters, hence the circular polarisation behaviour in this burst is not surprising in itself but provides additional evidence that repeating FRB progenitors likely reside in a more complex magneto-ionic environment.

Using the FR-GFR model, we find that the RM of the discovery burst is 27.7$_{-5.1}^{+5.2}$ \RMunits,  less extreme than the previously reported RM of $-3163$ \RMunits\, \citep{Price_2018}.  The sign of the RM is opposite, suggesting there has only been one observed flip in the sign of the burst between the discovery of the burst from the Parkes/Murriyang multibeam receiver and additional follow-up from Parkes/Murriyang UWL; between May 2021 and June 2022. 
The QUV fit for the data shown in Figure \ref{fig:180301A_QUV_fit} shows good agreement 
with the Stokes U and V. However, we see a relatively larger residual 
with Stokes Q. We attribute this to the differential gain in the two linearly polarisation feeds from the multibeam receiver. The Stokes-Q parameter with linear feeds is a differential parameter, compared to the Stokes U and V, which are cross correlation parameters. Hence, Stokes-Q is less immune to variations in differential gain.

Assuming the surrounding media to be dominated by relativistic gas plasma, we would expect the spectral index $\alpha=3$. From \cite{Kennett}, the GRM or relativistic rotation measure (RRM) for this case can be defined by GRM = $10^9$LB$^4$ \RMalpha, where B is the magnetic field in Gauss and L is the distance scale in parsec. If we set $\alpha=3$, we estimate the GRM to be 5121.84$^{+24.45}_{-25.59}$ \RMalphathree. Assuming a length scale of $\sim$1 pc (the fiducial scale used in \cite{Kennett}), the magnetic field strength for the media is $\sim$2.2 mG. This is comparable to measurements reported by \cite{Johnston_2005} for the 2004 periastron passage of B1259-63 system, albeit in a much more compact environment. 

In the recent observation of FRB~20180301A from Parkes/Murriyang UWL \citep{Pravir_2023}, no burst was observed to have circular polarisation $>15\%$, and there was no evidence for frequency-dependent circular polarisation\footnote{If there was a change in sign in $V$ it would be possible to have frequency dependent circular polarisation with small band-averaged total circular polarisation.}. This would indicate that the circular polarisation could be a transient characteristic due to changing circumburst environment of the FRB progenitor, similar to the behaviour seen in the FRB 20201124A, where significant frequency dependent circular polarisation was only seen in some of the bursts during its active phase in April 2021 \citep{Pravir_20201124A_MNRAS}. Additionally, observational bias could lead to missing bursts showing significant circular polarisation fraction during this transient phase. The FRB~20180301A provides a unique window into studying the progenitor environment of FRBs using its complex polarimetric properties. Hence, a high cadence monitoring of this source is essential to confirm any periodic polarimetric behaviour in FRB~20180301A.


\section{Conclusions}
\label{sec:Conclusisons}
We use GFR to model the Stokes-V behaviour of FRB~20180301A. We extend the RM estimation method using QU-fit described in \cite{Bannister_2019} and \cite{Lower_2021} to QUV-fit for GFR modelling and derive a combined RM-GFR estimation method. We verify this method using known pulsars J1644$-$4559 and J0835$-$4510. 

We use arcsec localisation from \cite{Bhandari_2022} and the pointing information from the observation to conclude that the burst was detected within the FWHP of the multibeam receiver, with a radial offset of $\sim$7' from the beam 3 centre. The precise localisation information suggests that the frequency-dependent Stokes-V burst behavior was likely non-instrumental. The combined estimation of the RM and GRM suggests a less extreme RM of $\sim$27 \RMunits, relative to the previously reported RM of -3163 \RMunits \citep{Price_2018}. Our results provide another example in an ever-growing subset of FRBs showing signature of passage through relativistic plasma. This confirms another promising way to use FRBs as probes of fundamental physics.  

\section*{Acknowledgements}

PAU and RMS acknowledges support through Australia Research Council Future Fellowship FT190100155. RMS also acknowledges support through Australian Research Council Discovery Project DP220102305. 
We acknowledge the Wiradjuri people as the Traditional Owners of the Observatory site. Murriyang, the Parkes radio telescope, is part of the Australia Telescope National Facility (\url{https://ror.org/05qajvd42}) which is funded by the Australian Government for operation as a National Facility managed by CSIRO. This work was performed on the OzSTAR national facility at Swinburne University of Technology. This work makes use of OzSTAR supercomputing facility. The OzSTAR program receives funding in part from the Astronomy National Collaborative Research Infrastructure Strategy (NCRIS) allocation provided by the Australian Government, and from the Victorian Higher Education State Investment Fund (VHESIF) provided by the Victorian Government.

This work makes use of \code{BILBY} (Ashton et al. 2019), \code{MATPLOTLIB} (Hunter 2007) and \code{NUMPY} (Harris et al. 2020) software packages. 

\section*{Data Availability}

This work used publicly available data. The codebase used for the GFR analysis is available on \hyperlink{https://github.com/pavanuttarkar/GFR\_codebase}{https://github.com/pavanuttarkar/GFR\_codebase}.



\bibliographystyle{mnras}
\bibliography{example} 




\appendix

\section{Circular polarisation likelihood equation}

The likelihood function for the GFR model described by Equation \ref{eq:Circular_model} can be described by  

\begin{equation}
\begin{split}
 &\textit{L}(P(\lambda)| P_M(\lambda) , \sigma) = 
 \frac{1}{\sigma_Q\sqrt{2\pi}}\exp{\frac{-(Q-Q_m)^2}{2\sigma_Q^2}} \\
 &\frac{1}{\sigma_U\sqrt{2\pi}}\exp{\frac{-(U-U_m)^2}{2\sigma_U^2}} 
 \frac{1}{\sigma_V\sqrt{2\pi}}\exp{\frac{-(V-V_m)^2}{2\sigma_V^2}},
\end{split}
\end{equation}

where $P(\lambda)$ is the polarised flux across wavelength $\lambda$, $P_M(\lambda)$ is the model polarised flux across wavelength, Q, U, and V are the Stokes components, $Q_m$, $U_m$, and $V_m$ are the model Stokes parameters, $\sigma_Q$, $\sigma_U$, and $\sigma_V$ are the rms value of the individual Stokes components. Using model parameters from Equation \ref{eq:Stokes} the above Equation simplifies to 

\begin{equation}
\begin{split}
    &\textit{L}(P(\lambda)| P_M(\lambda) , \sigma) = \\
    &\frac{1}{(2\pi)^{3/2}\sigma^3} \exp{-\frac{1}{2\sigma^2}(Q^2+U^2+V^2)} \\
    &\exp{\Biggr[\frac{1}{2\sigma^2}\left[P^2-2P(QQ_m+UU_m+VV_m)\right]\Biggr]},
    \end{split}
\label{eq:Stokes_reduced_sigma}
\end{equation}

where P is the total polarised flux, following \cite{Bannister_2019} we integrated the total polarised from Equation \ref{eq:Stokes_reduced_sigma}. We also note that \cite{Bannister_2019} used integration from -$\infty$ to $\infty$ which induces bi-modality in the likelihood distribution. We correct for this by integrating the above Equation from 0 to $\infty$, to avoid bi-modality

\begin{equation}
\begin{split}
    &\textit{L}(P(\lambda)| P_M(\lambda) , \sigma) = \\
    &\frac{1}{(2\pi)^{3/2}\sigma^3} \exp{\frac{-1}{2\sigma^2}(Q^2+U^2+V^2)} \\
    &\int_{0}^{\infty}\exp{\Biggr[\frac{1}{2\sigma^2}\left[P^2-2P(QQ_m+UU_m+VV_m)\right]\Biggr]} dP.
\end{split}
\label{eq:Stokes_reduced_sigma_integral}
\end{equation}

\begin{figure*}
        \subcaptionbox[width=2.28\columnwidth \label{fig:180301A_Posterior}]{}
	{\includegraphics[width=2.28\columnwidth]{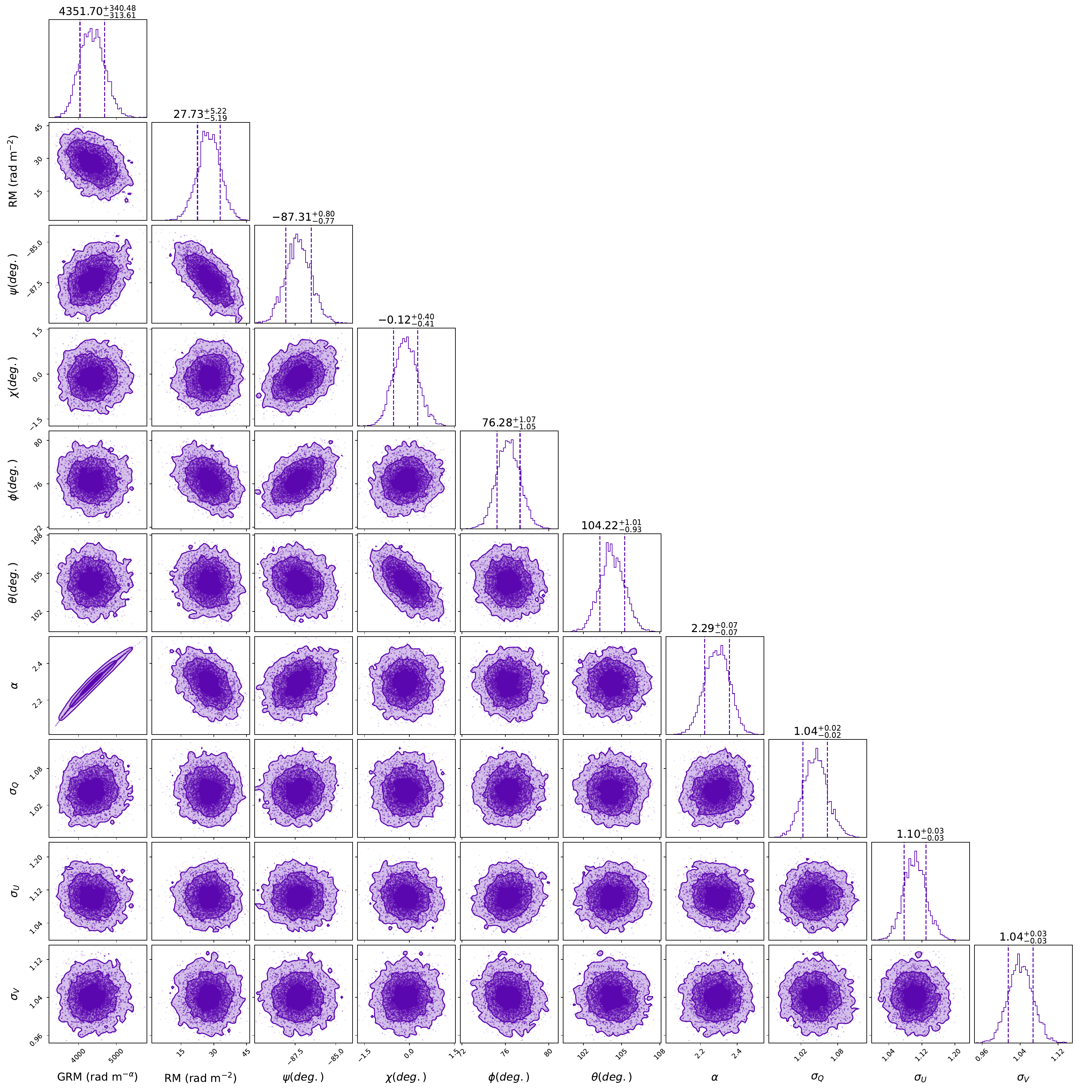}}
        
         \caption{The posterior probability distribution of the FR-GFR model for FRB~20180301A. The modelled posterior distribution is from the FR-GFR model described by Equation \protect\ref{eq:posterior_final}. The estimated parameters from the posterior distribution are listed in Table \protect\ref{tab:180301A_tab}. We use uniform priors for all the parameters shown in the posterior probability distribution. The dashed lines shows the 1-$\sigma$ confidence interval for all the parameters. The three contours represent 1-$\sigma$, 2-$\sigma$, and 3-$\sigma$ confidence intervals.}
         \label{fig:180301A_posteiors}
         \vspace{2cm}
\end{figure*}

\begin{figure*}
        \subcaptionbox[width=1\columnwidth \label{subfig:B0833_QVU}]{}
	{\includegraphics[width=1\columnwidth]{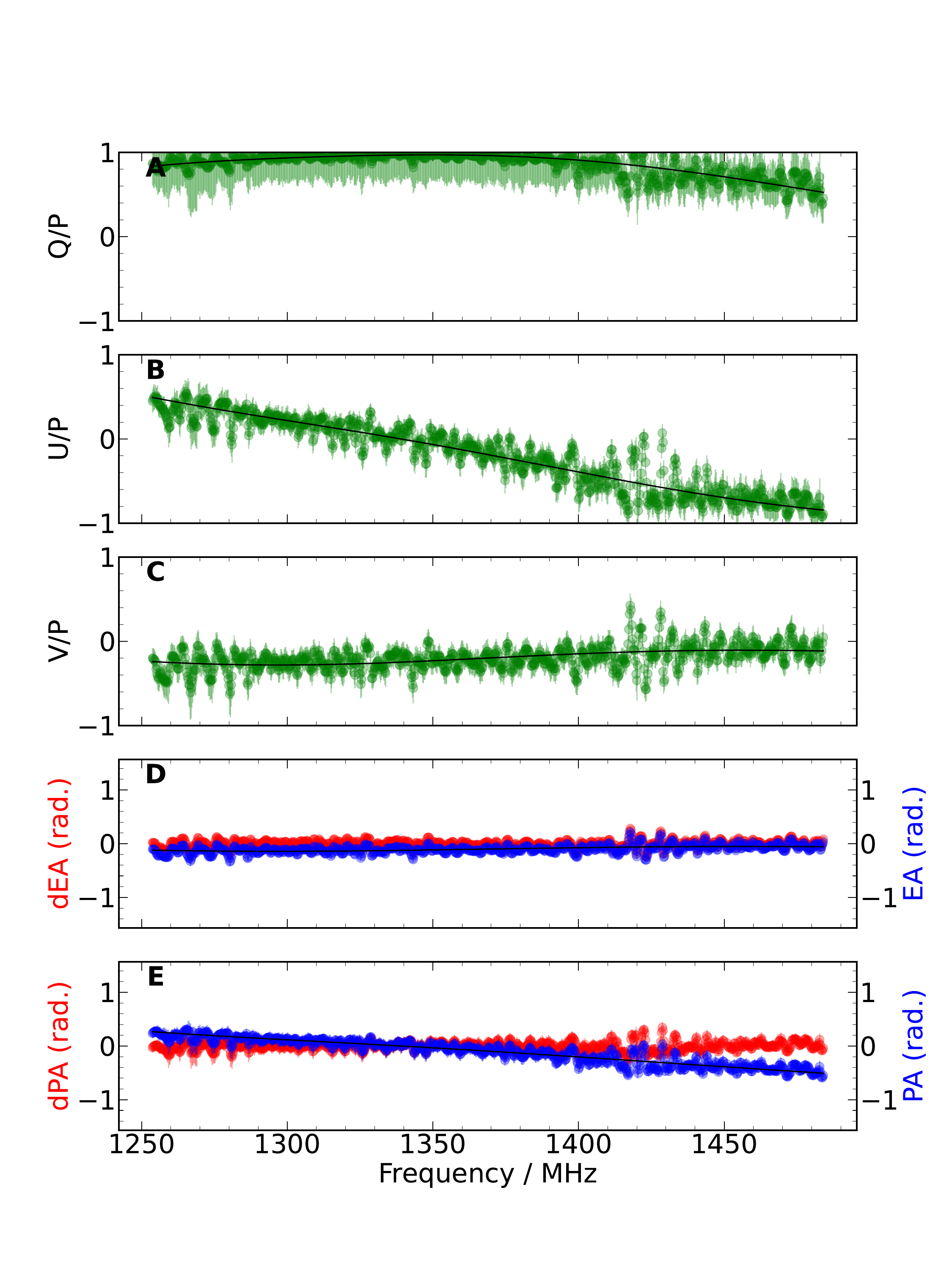}}
        \subcaptionbox[width=1\columnwidth \label{J1644_QVU}]{}
	{\includegraphics[width=1\columnwidth]{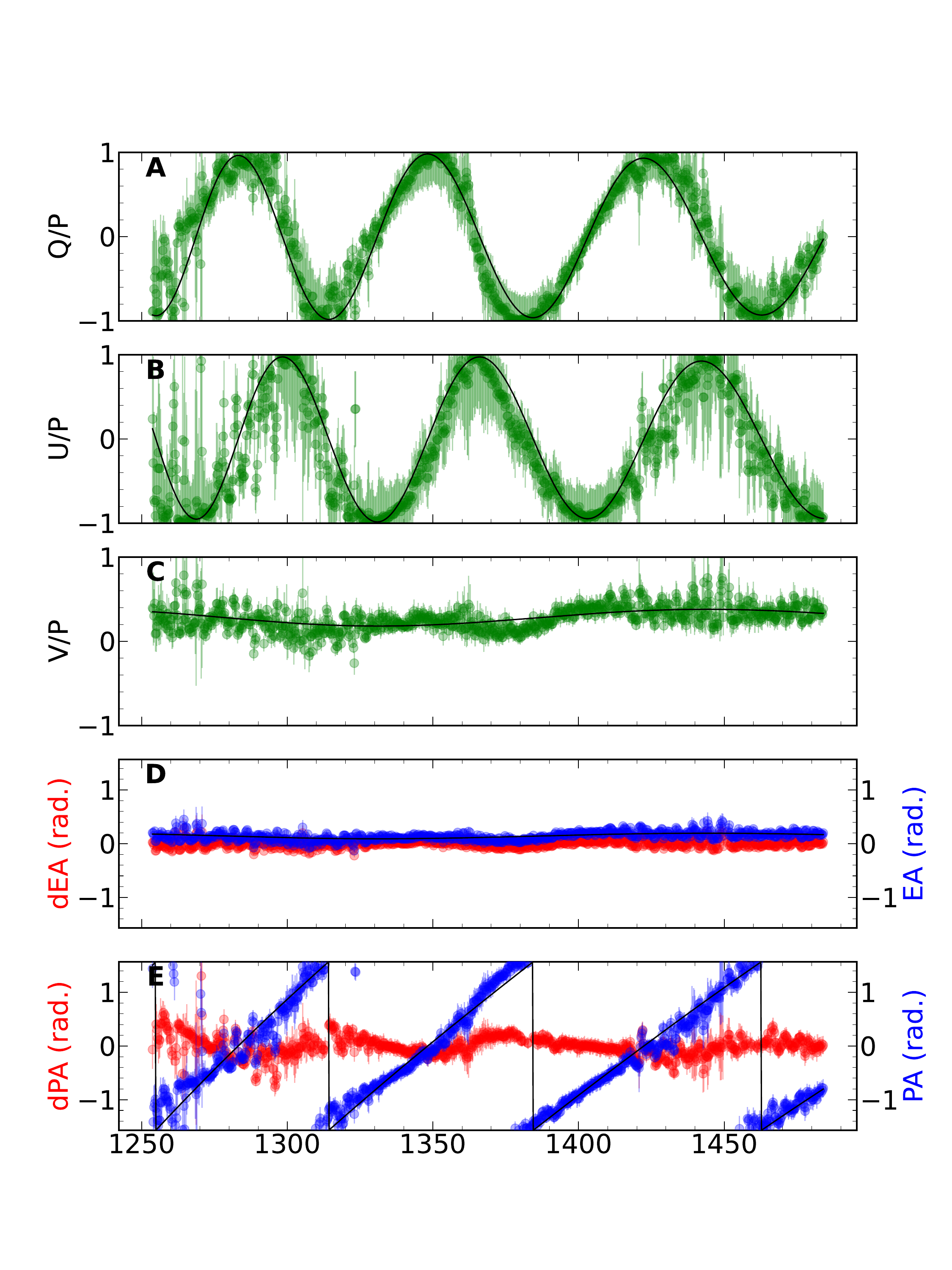}}
         \caption{The QUV fit for pulsars J0835$-$4510 and J1644$-$4559. Figures (a) and (b) show the model fit derived for the polarisation calibrated dataset for pulsars J0835$-$4510 and J1644$-$4559, respectively. The model fits are derived using the FR-GFR model. Stokes Q, U, and V fit are shown in Panels A, B, and C, respectively, for each figure. We use archival data (project ID P595) from the Parkes/Murriyang radio telescope. The data for the time averaged on-pulse region are shown in green coloured points, and the solid black line shows the model fit to the data. The EA and PA angles derived from the data are shown in blue points in Panels D and E, respectively, for Figures (a) and (b). The model fit for the EA and PA angles are shown in black solid lines. The residual after subtracting the data from the model is shown in red points for EA and PA angles in Panels D and E, respectively.}
         \label{fig:pulsar_QUV_fit}
\end{figure*}

\begin{figure*}
	{\includegraphics[width=2\columnwidth]{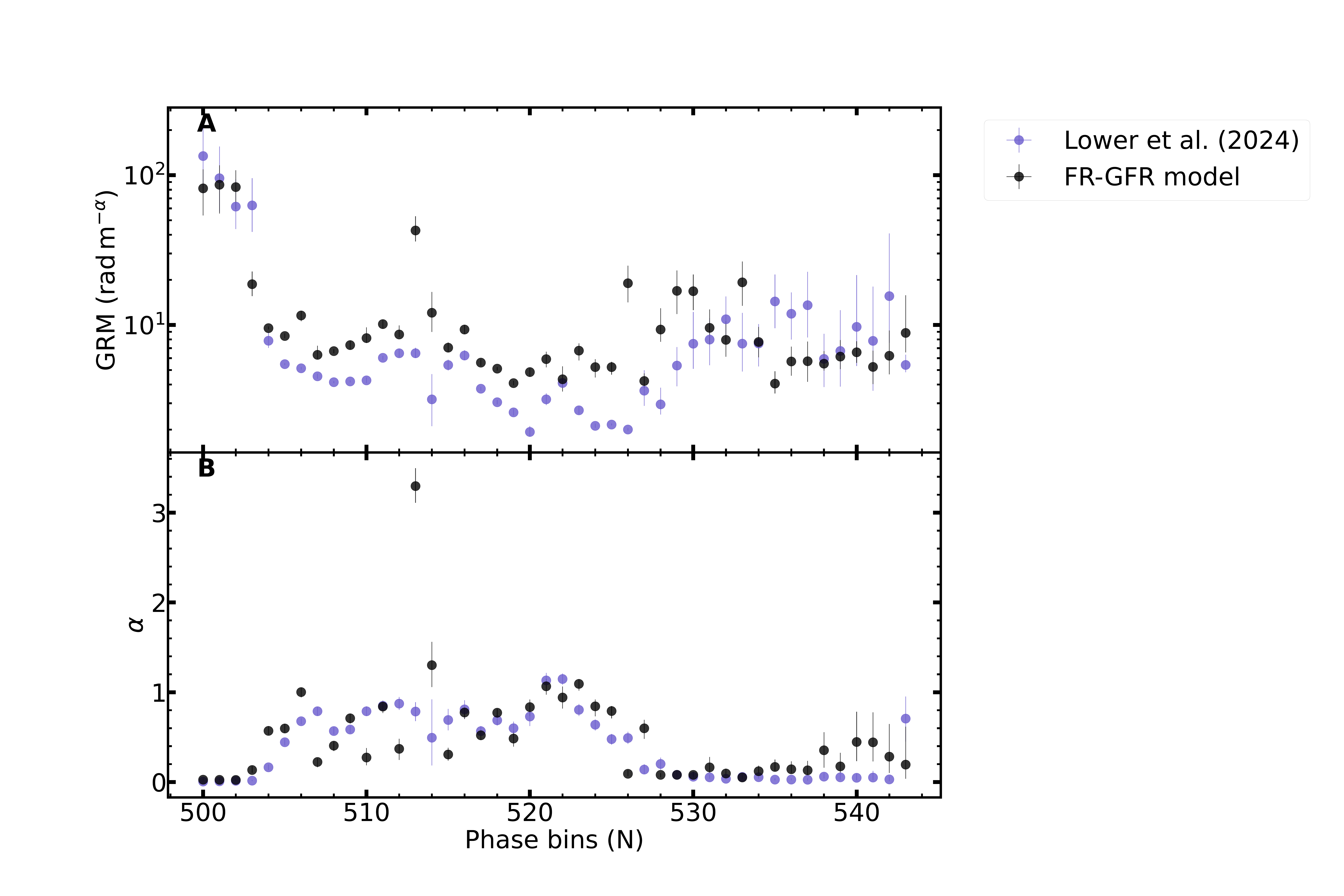}}
         \caption{The FR-GFR model comparison with results from \protect\cite{Lower_2023} for magnetar J1810$-$197. The panels A and B show the GRM and $\alpha$ values recovered for J1810$-$197 for different pulse phase bins. The black and blue points show the recovered values from \protect\cite{Lower_2023} and the FR-GFR model. The errorbars on individual points show 1-$\sigma$ error range of recovered values.}
         \label{fig:J1810_verififcation}
\end{figure*}

Further, Equation \ref{eq:Stokes_reduced_sigma_integral} reduces to Equation \ref{eq:posterior_final}. The posterior probability distributions obtained from the likelihood Equation \ref{eq:posterior_final} for test pulsars J1644$-$4559 and J0835$-$4510 are shown in Figure \ref{fig:pulsar_posteiors}. The posterior probablility distribution for FRB 20180301A is shown in Figure \ref{fig:180301A_posteiors}. The model fit to pulsars J1644$-$4559 and J0835$-$4510 is shown in Figure \ref{fig:pulsar_QUV_fit}. 
A comparison of the GFR parameters for the magnetar XTE J1810$-$197 can be found in Figure \ref{fig:J1810_verififcation}.

\begin{figure*}
        \subcaptionbox[width=2\columnwidth \label{fig:B0833_Posterior}]{}
	{\includegraphics[width=2\columnwidth]   {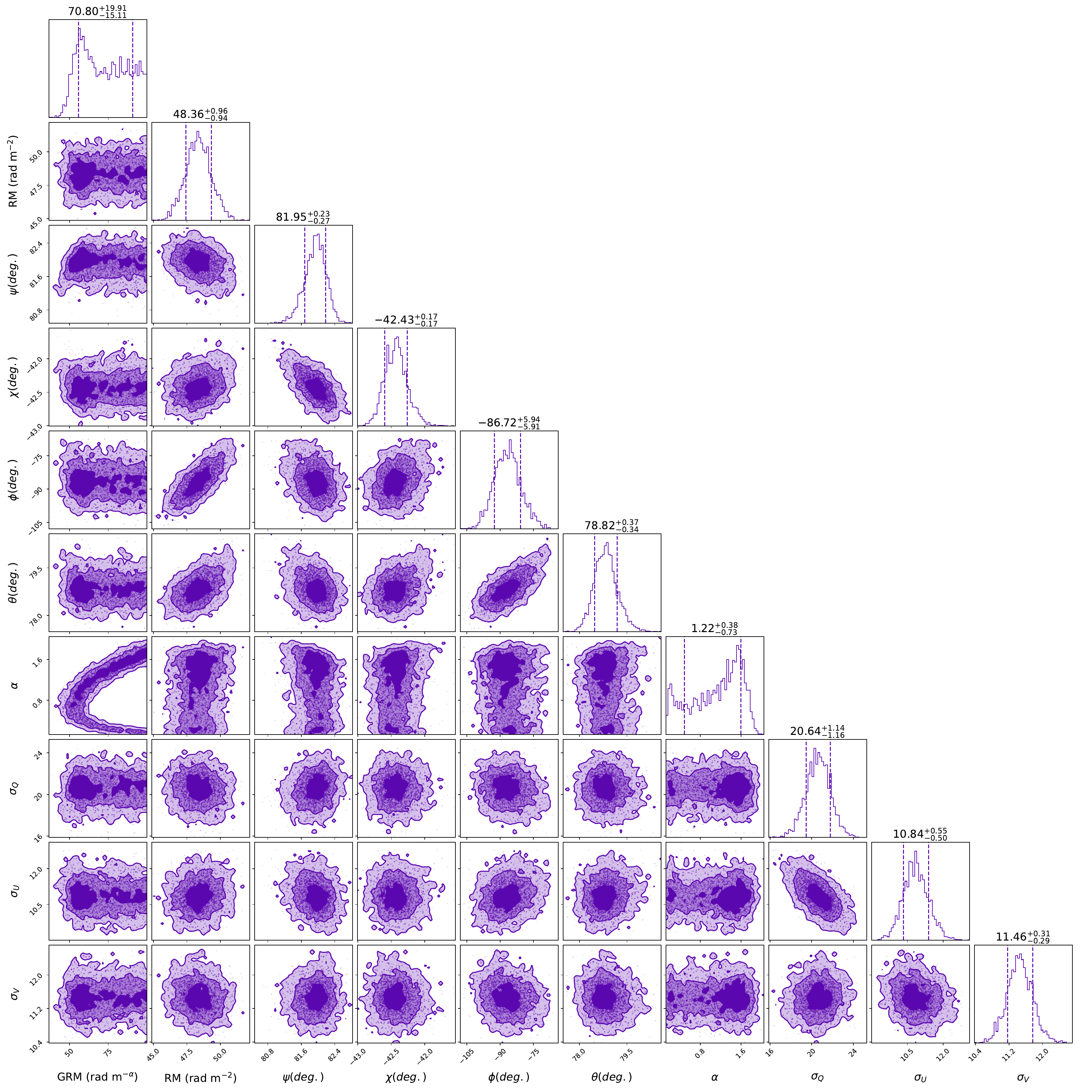}}
\end{figure*}
\begin{figure*}
    \ContinuedFloat  
            \subcaptionbox[width=2\columnwidth \label{fig:J1644_Posterior}]{}
	{\includegraphics[width=2\columnwidth]   {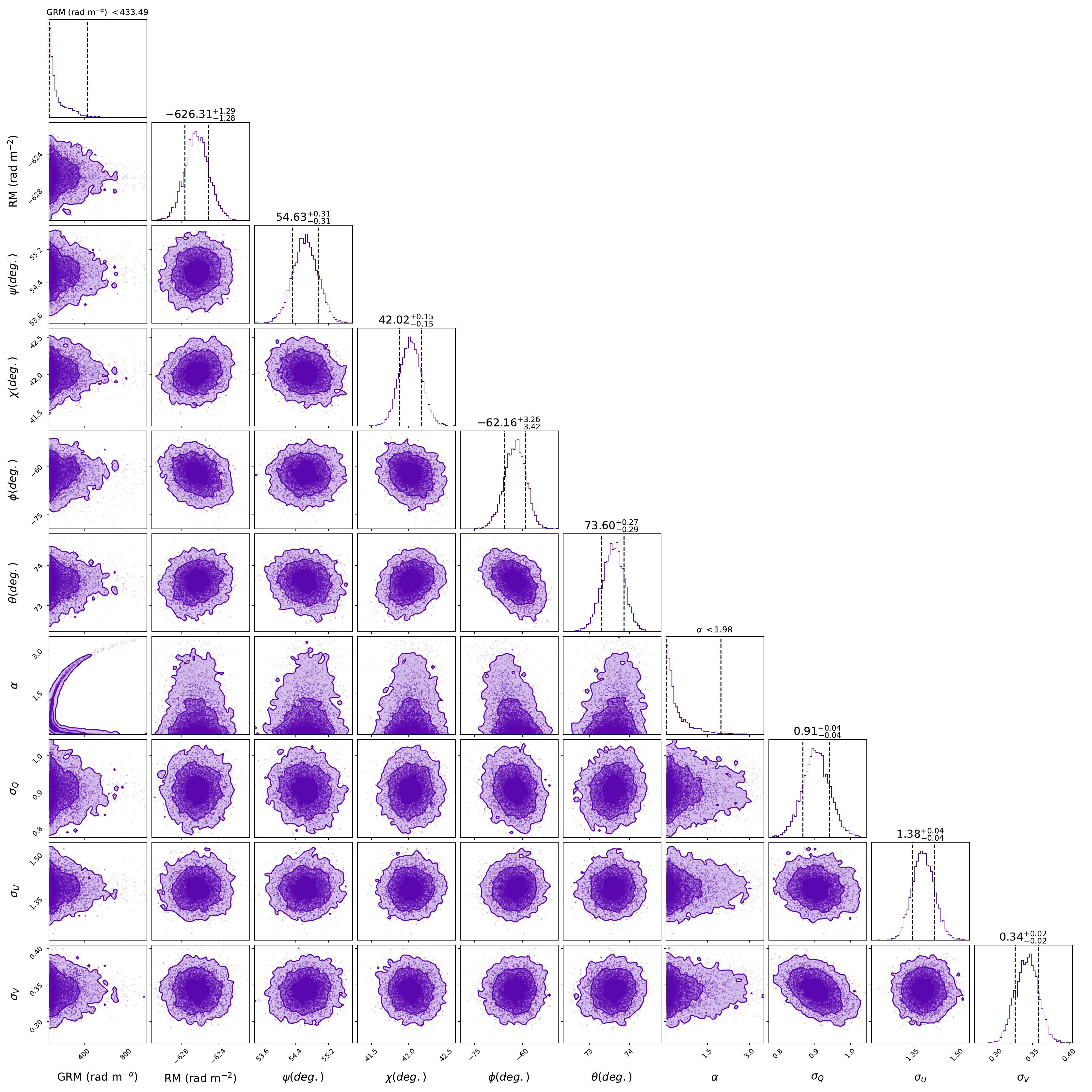}}
         
         \caption{The posterior distribution for the test pulsars J1644-4559 and J0835-4510. We use archival multibeam data for the verification of the model. The corner plots shows the posterior parameters as listed in Table \ref{tab:comparision_table}. We use uniform priors for all the parameters. The dashed lines show 1-$\sigma$ confidence interval for all the parameters. The inner and outer contours show 1-$\sigma$ and 3-$\sigma$ intervals.}
         \label{fig:pulsar_posteiors}
\end{figure*}

\bsp	
\label{lastpage}
\end{document}